\documentclass[twocolumn]{aastex631}

\usepackage{graphicx}	
\usepackage{amsmath}	
\usepackage{amssymb}	
\usepackage{color}
\usepackage{ulem}

\shorttitle{Narrowing down FDM Mass with Ultra-faint dwarfs}
\shortauthors{Hayashi et al.}
\graphicspath{{./}{figures/}}

\begin{document}

\title{Narrowing the mass range of Fuzzy Dark Matter with Ultra-faint Dwarfs}

\author[0000-0002-8758-8139]{Kohei Hayashi}
\correspondingauthor{Kohei Hayashi}
\affiliation{Astronomical Institute, Tohoku University, Sendai, Miyagi, 980-8578, Japan}
\email{k.hayasi@astr.tohoku.ac.jp}

\author[0000-0002-5032-8368]{Elisa G. M. Ferreira}
\affiliation{Max Planck Institute for Astrophysics, Karl-Schwarzschild-Str. 1, 85748 Garching, Germany}
\email{elisagmf@mpa-garching.mpg.de}

\author[0000-0001-9053-6922]{Hei Yin Jowett Chan}
\affiliation{Astronomical Institute, Tohoku University, Sendai, Miyagi, 980-8578, Japan}
\email{jchan@astr.tohoku.ac.jp}


\begin{abstract}
       Fuzzy dark matter~(FDM) is an attractive dark matter candidate motivated by small scale problems in astrophysics and with a rich phenomenology on those scales.  We scrutinize the FDM model, more specifically the mass of the FDM particle, through a dynamical analysis for the Galactic ultra-faint dwarf (UFD) galaxies. We use a sample of 18 UFDs to place the strongest constraints to date on the mass of the FDM particle, updating on previous bounds using a subset of the sample used here. 
       We find that most of the sample UFDs prefer a FDM particle mass heavier than $10^{-21}$eV. In particular, Segue~1 provides the strongest constraint, with $m_\psi=1.1^{+8.3}_{-0.7}\times10^{-19}\mathrm{eV}$. The constraints found here are the first that are compatible with various other independent cosmological and astrophysical bounds found in the literature, in particular with the latest bounds using the Lyman-$\alpha$ forest. We also find that the constraints obtained in this work are not compatible with the bounds from luminous dwarf galaxies, as already pointed out in the previous work using UFDs. This could indicate that although a viable dark matter model, it might be challenging for the FDM model to solve the small scale problems. 
\end{abstract}




\section{Introduction}

The Fuzzy Dark Matter (FDM) model~\citep[e.g.,][]{Hu:2000ke} recently emerged as an alternative that can solve the small-scale challenges to the Cold Dark Matter (CDM) model, while preserving the behaviour of CDM on large scales. Due to its ultra-light mass and bosonic nature, FDM exhibits a wave-like behaviour on Galactic scales, that leads to cosmological and astrophysical consequences. The effects of these on different observables allows to constrain the one and only parameter, the particle mass $m_\psi$ of the FDM model~\citep[for reviews on these effects see][]{Ferreira:2020fam,Hui:2021tkt}.

The proposed mass range for which the FDM produces a solitonic core in the interior of galaxies, thus addressing the ``core-cusp'' problem, is of the order of $m_\psi \sim 10^{-22}$~eV~\citep{Hui:2016ltb}. However, the current observations are pushing the lower bound to heavier particle masses. 
The suppression in the matter power spectrum induced by FDM can be measured by the Lyman-$\alpha$ forest observations, and has put a lower bound of $m_\psi \ge 10^{-20}$ eV \citep{Rogers2020}.
These bounds are in tension with the previously expected canonical value of $m_\psi$ required to solve the small-scale problem.

The dwarf spheroidal galaxies (dSphs), as a dark matter dominated system, are the promising astrophysical probes of the nature of dark matter. 
The stellar kinematic data of the dSphs therefore enable us to place constraints on the particle mass based on the dynamical analysis~\citep{Schive2014,Chen:2016unw,Gonzales-Morales:2016mkl, Hayashi2020}.
However, several groups have suggested that dark matter density profiles in the luminous dSphs could be affected by baryonic physics~\citep[e.g.,][]{Read:2018fxs,Hayashi:2020jze}.
For the ultra faint dwarf (UFD) galaxies, which host much smaller stellar masses than the luminous ones, the impact of baryonic feedback on their inner dark matter densities can be negligible~\citep[e.g.,][]{2020MNRAS.497.2393L} and thus the UFDs are ideal targets to derive reliable constraints on the FDM model.
Nevertheless, only a few have studied the FDM model by UFDs. 
Using the measurement of half-light masses of two UFDs, Draco II and Triangulum II, a mass estimate of $m_{\psi} \sim 3.7-5.6 \times 10^{-22}$ eV was obtained by~\citet{Calabrese:2016hmp}. In~\cite{Safarzadeh:2019sre}, using Milky Way UFDs, a bound on $m_\psi$ was obtained, $m_\psi > 10^{-21}$~eV. This bound was compared with the bounds from the analysis of luminous dSphs, showing that those are in tension with each other.
As more stellar kinematic data of UFDs become available, it is essential to perform a full Jeans analysis to update the constraint on $m_\psi$.

In this letter, we perform the Jeans analysis of the stellar kinematic data of 18 UFDs to constrain $m_\psi$. We show that these systems prefer $m_\psi$ that is higher than the canonical expectation of the FDM. We compare our mass constraints to the most relevant bounds that exist in the literature and found that they are compatible within uncertainties. Among the UFDs, Segue 1 provides the most strongest constraint, which challenges the proposed canonical mass range of the FDM model with $m_\psi \sim 10^{-22} - 10^{-21}$ eV. 

\section{Models}

To constrain $m_\psi$, we adopt here the spherical Jeans equation, which relates a stellar phase-space distribution to a dark matter halo mass distribution. 
For a spherically symmetric system in dynamical equilibrium, this is given by~\cite{2008gady.book.....B}:
\begin{equation}
    \frac{\partial[\nu(r)\sigma^2_r(r)]}{\partial r} + \frac{2\nu(r)\beta_{\mathrm{ani}}(r)\sigma^2_r(r)}{r} = -\nu(r)\frac{GM(r)}{r^2}\,,
    \label{eq:Jeans}
\end{equation}
where $r$ denotes the radius from the center of a system, $\nu(r)$ is the three dimensional stellar density distribution, $G$ is the gravitational constant, and $M(r)$ is the dark matter mass distribution. 
The stellar velocity ellipsoid defined by $(\sigma_r,\sigma_{\theta},\sigma_{\phi})$ is equationed with spherical coordinates.
Since $\sigma_{\theta}=\sigma_{\phi}$ for spherical symmetry, the stellar velocity anisotropy is written as $\beta_{\mathrm{ani}}(r)\equiv1-\sigma^2_{\theta}(r)/\sigma^2_{r}(r)$.
Here, we adopt a general and realistic stellar anisotropy model proposed by \cite{Baes:2007tx}:
\begin{equation}
\label{eq:anisotropy}
\beta_{\mathrm{ani}}(r) = \frac{\beta_0+\beta_{\infty}(r/r_{\beta})^{\eta}}{1+(r/r_{\beta})^{\eta}}\,,
\end{equation}
where $\beta_0$ and $\beta_{\infty}$ are the inner and outer anisotropy parameters. The spatial dependence of $\beta_{\mathrm{ani}}$ is characterized by the transition sharpness $\eta$ and radius $r_{\beta}$.

To compare observed line-of-sight velocity dispersion profiles with the model, we project the radial dispersion profile from equation~(\ref{eq:Jeans}) into the line-of-sight direction:
\begin{equation}
\label{eq:losdisp}
\Sigma(R)\sigma_{\mathrm{ los}}^2(R)=2\int_R^\infty\!\! dr \Bigl(1-\beta_{\mathrm{ani}}(r)\frac{R^2}{r^2}\Bigr)\frac{\nu(r)\sigma_r^2(r)}{\sqrt{1-R^2/r^2}}\,,
\end{equation}
where $R$ is the projected radius, and $\Sigma(R)$ is the projected stellar density profile derived from the intrinsic stellar density $\nu(r)$, and $\sigma_{\mathrm{los}}$ is the line-of-sight velocity dispersion.
In this work, we assume the Plummer profile for the stellar density profile~\citep{Plummer:1911zza}, $\Sigma(R)=(\pi r^2_{\mathrm{half}})^{-1} [1+R^2/r^2_{\mathrm{half}}]^{-2}$ where $r_{\mathrm{half}}$ is the projected half-light radius.

FDM can form a soliton core in the central parts of a galaxy, which corresponds to the ground state solution of the Schr\"odinger-Poisson equation.
Owing to high-resolution FDM simulations~\citep[e.g.,][]{Schive2014}, the radial profile of the soliton core can be written analytically: 
\begin{equation}
    \rho_{\rm soliton}(r) = \frac{\rho_c}{[1+0.091(r/r_c)^2]^8}\,,
    \label{eq:soliton_dens}
\end{equation}
where $r_c$ is the soliton core radius and $\rho_c$ is the central density given by
\begin{equation}
    \rho_{c} = 1.9\times10^{12} \Bigl(\frac{m_{\psi}}{10^{-23}\ {\rm eV}}\Bigr)^{-2} \Bigl(\frac{r_c}{{\rm pc}}\Bigr)^{-4} \ [M_{\odot}\ {\rm pc}^{-3}]\,.
\end{equation}
The mass profile, $M_{\rm soliton}(r)=\int^{r}_{0}4\pi\,s^2\rho(s)ds$ can be also calculated analytically~\citep{Chen:2016unw}.
The simulations also predicted the scaling relation between a soliton core radius, particle mass, and dark halo mass,
\begin{equation}
     r_c \simeq 1600\left(\frac{m_{\psi}}{10^{-23} \mathrm{eV}}\right)^{-1} \left(\frac{M_{\mathrm200}}{10^{12}M_{\odot}}\right)^{-1/3}\ \mathrm{pc}\,,
     \label{eq:scaling}
\end{equation}
where $M_\mathrm{200}$ is the enclosed mass within $r_\mathrm{200}$ in which the spherical overdensity is 200 times the critical density of the universe.
We thus determine the soliton core radius $r_c$, given $m_\psi$ and $M_{\mathrm{200}}$ as free parameters.

Beyond the core radius, the halo profile is akin to a Navarro-Frenk-White~(NFW) profile~\citep{Navarro:1996gj},
\begin{equation}
    \rho_{\mathrm{NFW}}(r) = \frac{\rho_{s}}{(r/r_s)(1+r/r_s)^2}\,.
\end{equation}
To transition from the central soliton core to the outer NFW halo, we impose a density continuity condition at the radius $r_\epsilon$,
\begin{equation}
    \frac{\rho_{s}}{(r_\epsilon/r_s)(1+r_\epsilon/r_s)^2}=\frac{\rho_{c}}{[1+0.091(r_\epsilon/r_c)^2]^8} = \epsilon\rho_c\,,
\end{equation}
and $r_\epsilon$ can be derived as $r_\epsilon=(0.091)^{-1/2}r_c(\epsilon^{-1/8}-1)^{1/2}$.
Thus, when $\epsilon$ and $r_s$ are given, $\rho_s$ can be determined.
According to the numerical simulations~\citep{Schive2014,Mocz:2018ium}, $r_\epsilon$ should be larger than $3r_c$. 

The halo with a heavier $m_\psi$ has a very small core and most of the halo is described by an NFW profile. Since a large part of the halo has a similar behaviour to a CDM halo, we impose the concentration-mass relation of the NFW halos predicted by CDM simulations:
\begin{eqnarray}
C_{200}(M_{200},x_{\mathrm{sub}}) = && c_0 \left[1+\sum^{3}_{i=1} \left[a_i \log_{10}\left(\frac{M_{200}}{10^8 h^{-1}M_{\odot}}\right)\right]^i \right] \nonumber \\
&& \times \left[1+b\log_{10}(x_{\mathrm{sub}})\right]. \label{eq:c200}
\end{eqnarray}
Here, we utilize $c_0=19.9$, $a_i=\{-0.195,0.089,0.089\}$, and $b=-0.54$, which are the best-fit parameters for the concentration-mass relation~\citep{Moline:2016pbm}.
The subhalo distance from the center of a host halo divided by $r_{200}$ of the host halo is given by $x_{\mathrm{sub}}\equiv r_{\mathrm{sub}}/r_{200,\mathrm{host}}$.
The $r_{200}$ of the Milky Way halo, $r_{200,\mathrm{MW}}$, has a large uncertainty $r_{200}\simeq210\pm50$~kpc.
However, the error of $r_{200,\mathrm{MW}}$ may not have impact on the concentration parameter.
Therefore, we adopt $r_{200,\mathrm{MW}}=210$~kpc in this work~\footnote{We performed the same MCMC analyses for the case of $r_{200,\mathrm{MW}}=160$ and $260$~kpc and obtained $m_\psi=1.0^{+8.6}_{-0.6}\times10^{-19}$~eV and $1.1^{+8.8}_{-0.7}\times10^{-19}$~eV, respectively. Thus, we confirmed the uncertainty on $r_{200,\mathrm{MW}}$ can be negligible.}.

\section{Data and Analysis}

To scrutinize FDM halos of the Milky Way UFDs, we select 18 galaxies~(Bo\"otes~I, Coma~Bernices, Canes~Venatici~I, Canes~Venatici~II, Eridanus~II, Grus~2, Hercules, Hydra~II, Leo~IV, Reticulum~II, Segue~1, Segue~2, Triangulum~II, Tucana~3, Tucana~4, Ursa~Major~I, Ursa~Major~II, and Willman~1).
They have more than 10 stellar kinematic data of their member stars.

The stellar structural parameters of these galaxies are taken from~\citet{2015ApJ...813..109D} and \citet{2018ApJ...860...66M}.
For the stellar kinematic sample analyzed in the present study, we use the published data from the original spectroscopic observation papers~\citep{2007ApJ...670..313S,2011ApJ...733...46S,2011AJ....142..128W,2013ApJ...770...16K,2015ApJ...810...56K,2015ApJ...808...95S,2017ApJ...838...83K,2017ApJ...838...11S,2020ApJ...892..137S,2021arXiv210100013J,Zoutendijk:2021kee}.
To identify the member stars, we adopt the methods in the above literature.
For the influence of unresolved binary stars on a stellar kinematics, several papers indicated that multi epoch observations can exclude binary candidates from stellar spectroscopic data and concluded that the presence of binaries is likely to have only mild influence on estimates of the velocity dispersion of UFDs. 
Thus, we ignore this effect.

Given the available observational data, we fit models for $\rho_{\mathrm{soliton}}(r) + \rho_{\mathrm{NFW}}(r)$ and $\beta_{\mathrm{ani}}(r)$ with the likelihood function $\log(\mathcal{L}_\mathrm{tot})=\log(\mathcal{L}_\mathrm{vel})+\log(\mathcal{L}_\mathrm{NFW})+\log(\mathcal{L}_{r_\mathrm{half}})$.
We assume that the line-of-sight velocity distribution is a Gaussian; thus the likelihood function coming from the stellar kinematics is written by
\begin{equation}
-2\log(\mathcal{L}_\mathrm{vel}) = \sum_{i}\left[\frac{(v_i-\langle v \rangle)^2}{\sigma^2_i} + \log(2\pi\sigma^2_i)\right]\,,
\label{eq:likeli_1}
\end{equation}
where $v_i$ and $R_i$ are the line-of-sight velocity and the projected radius from the center of the galaxy of the $i^{\mathrm th}$ star in the kinematic sample. The averaged line-of-sight velocity of the member stars $\langle v \rangle$ is a nuisance parameter.
The dispersion $\sigma^2_i$ can be written by the measurement error $\delta_{v,i}$ and the intrinsic dispersion: $\sigma^2_i=\delta^2_{v,i}+\sigma^2_{\mathrm{los}}(R_i)$.

Meanwhile, we require the outer NFW halo to satisfy the concentration-mass relation; thus, the likelihood is given by 
\begin{equation}
    -2 \log({\cal L}_{\rm NFW}) = \frac{[\log_{10}(c_{200}) - \log_{10}(C_{200})]^2}{\sigma^2_{\rm CDM}}\,.
\end{equation}
We estimate $c_{200}$ from the parameters $(r_s,\rho_s,M_{200})$ and $C_{200}$ is the median subhalo concentration-mass relation in Eq.~\ref{eq:c200}
with $\sigma_{\rm CDM} = 0.13$~\citep{Moline:2016pbm}.

We also consider the uncertainties of the half-light radius ($r_{\mathrm{half}}$) of the Plummer profile as the following form: 
$-2\log(\mathcal{L}_{r_\mathrm{half}})=(r_\mathrm{half}-r_\mathrm{half,obs})^2/\delta r^2_\mathrm{half,obs}$,
where $r_\mathrm{half,obs}$ and $\delta r_\mathrm{half,obs}$ are the measured half-light radius and its error based on the photometric data.

Our model has 10 free parameters~$(m_{\psi}$, $M_{\mathrm{200}}$, $\epsilon$, $r_s$, $\beta_0$, $\beta_\infty$, $r_\beta$, $\eta$, $r_\mathrm{half}$ and $\langle v \rangle)$.
We adopt flat priors over the following ranges: 
$-3\leq\log_{10}(m_{\psi}/10^{-23}\mathrm{eV})\leq8$, 
$8\leq\log_{10}(M_{\mathrm{200}}/M_{\odot})\leq11$,
$-5\leq\log_{10}(\epsilon)\leq \log_{10}(0.5)$,
$0\leq\log_{10}(r_{s,\beta}/[\mathrm{pc}])\leq4$,
$1 \le \eta \le 10, ~0 \le 2^{\beta_{0(\infty)}} \le 1(2),
~0 \le r_\mathrm{half}/\mathrm{pc} \le 1000$, and $-1000\le \langle v \rangle/(\mathrm{km~s^{-1}}) \le +1000$.
For the range of $M_{\mathrm{200}}$, several numerical studies implied that UFD-sized galaxies~($L_\ast\sim10^{2.5-5.0}L\odot$) reside in dark matter halos of mass $M_{\mathrm{200}}\sim10^8-10^9M_\odot$~\citep[e.g.,][]{2015MNRAS.453.1305W}. To obtain conservative limits on $m_\psi$, we however adopt the heavier upper limit of this prior range.
We map the posterior distributions of these parameters using the public python package {\it emcee}~\citep{2013PASP..125..306F}.
For each galaxy, we set the sampler to 280 walkers and 6000 steps and removed 1000 steps as burn-in.

\section{Results and discussion}

\begin{figure}[t!]
    \centering
    \includegraphics[width=\columnwidth]{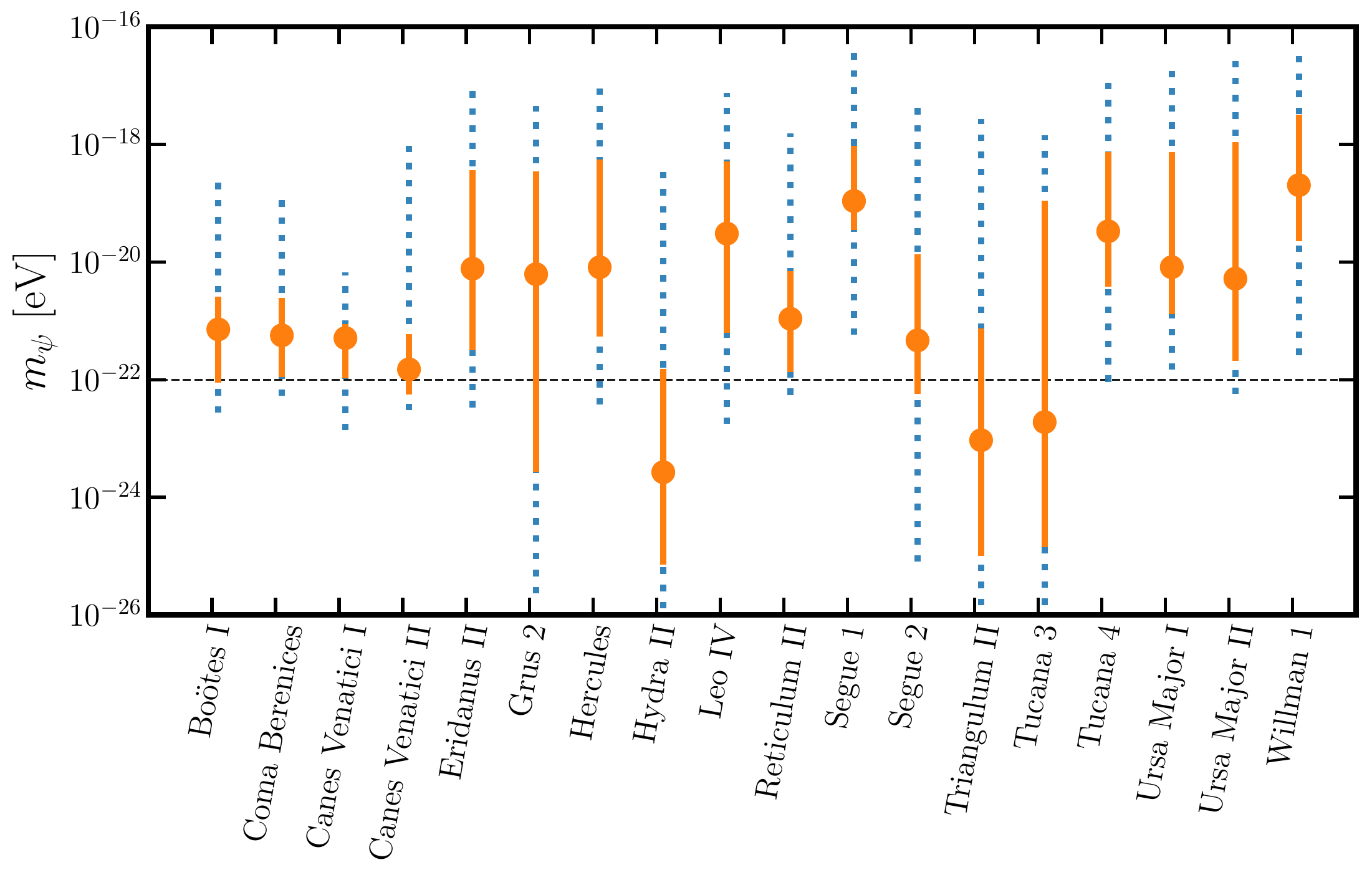}
    \caption{Estimated FDM mass, $m_{\psi}$, for 17 UFDs. The points show the median values of $m_{\psi}$, and orange solid and blue dotted errors are 1- and 2-$\sigma$ credible intervals. The horizontal dashed line corresponds to $m_{\psi}=10^{-22}~\mathrm{eV}$.}
    \label{fig:m_psi}
\end{figure}

In Fig.~\ref{fig:m_psi}, we show the estimated $m_\psi$, from all galaxies.
The median, 1- and 2-$\sigma$ credible intervals are computed from the posterior probability distribution functions~(PDFs).
We find that even though there surely exist large uncertainties on $m_\psi$ caused by a small sample size, the majority of the UFDs favor $m_\psi \gtrsim 10^{-22}~\mathrm{eV}$~(the horizontal dashed line in this figure), which was considered the key particle mass to resolve the core-cusp problem~\citep[e.g.,][]{Marsh:2015wka}.
In particular, the fitting result for Segue~1 shows the significantly large FDM mass, $m_\psi=1.1^{+8.3(+403)}_{-0.7(-1.1)}\times10^{-19}~\mathrm{eV}$ at 1- and 2-$\sigma$ credible intervals\,\footnote{
The estimated dark halo mass, $M_{200}$, for Segue~1 is $\log_{10}(M_{200}/M_\odot)=9.6^{+0.74}_{-0.68}$ at $1\sigma$ credible intervals.}. These results show that the soliton core in UFDs is very small, which confirms the existence of a large NFW outer halo around the solitonic core, supporting the profile used in the fit of this work.

\begin{figure}[t!]
    \centering
    \includegraphics[width=\columnwidth]{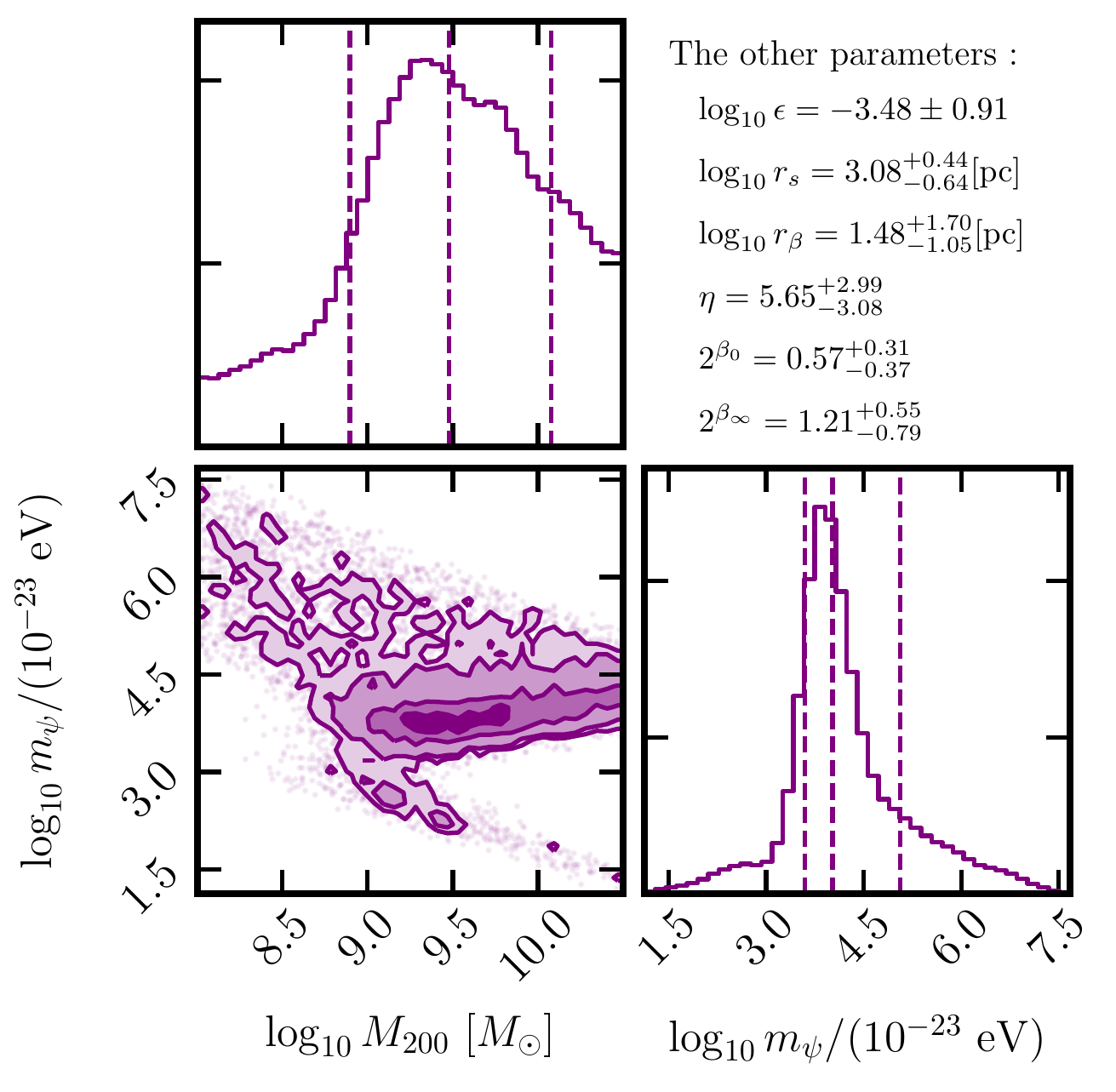}
    \caption{Posterior distributions of $M_{200}$ and $m_\psi$ for Segue~1. The dashed vertical lines in the 1D histograms are the median and 1$\sigma$ credible intervals.
    At the upper right corner, the other parameters~(the median and 1$\sigma$ credible intervals) are indicated.}
    \label{fig:PDFs}
\end{figure}

Fig.~\ref{fig:PDFs} shows the posterior PDFs of Segue~1. 
To focus on $m_\psi$, we show the posteriors of $M_{200}$ and $m_\psi$ only, while the other parameters are indicated at the upper-right corner. As shown in this figure, $m_\psi$ can be constrained {\it statistically} by our analysis.
Note that the constraint on $m_\psi$ from Segue~1 comes mainly from the kinematic sample in the inner part (especially within 10~pc) in our {\it unbinned} analysis. The unbinned analysis can trace the inner kinematic structures in the UFDs, while the binned one might smear out such information, and thus may not provide such a strong constraint on $m_\psi$ as the unbinned one.

%

\begin{figure}[t!]
    \centering
    \includegraphics[width=\columnwidth]{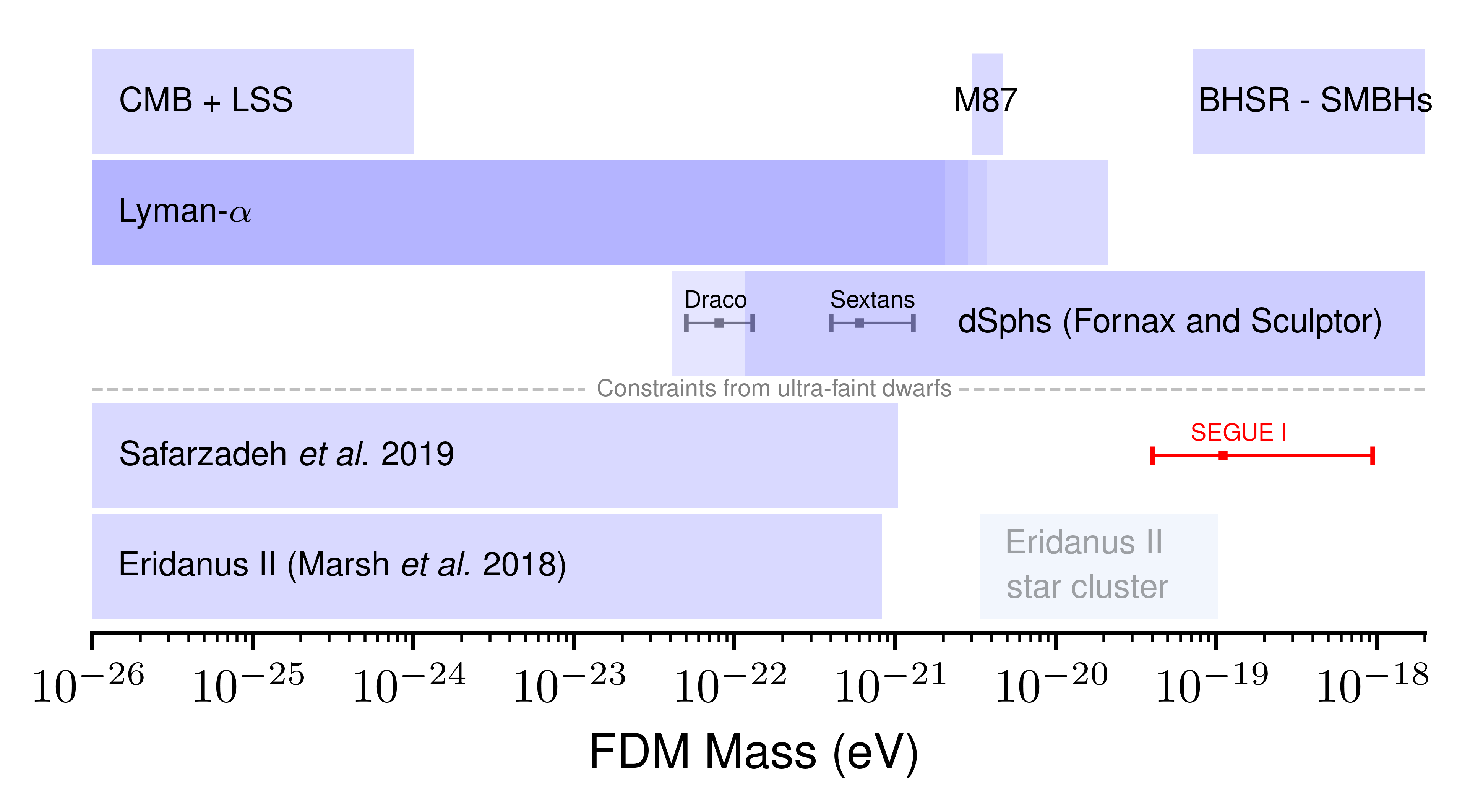}
    \caption{Summary of the most relevant constraints on the FDM mass to date. The blue shaded regions represent the excluded regions.
    We highlight in red the constraint for Segue~1 from this work. The first line presents the bounds from Cosmic Microwave Background~(CMB) and Large Scale Structure~(LSS) from~\cite{Hlozek:2014lca,Hlozek:2017zzf}, and those from black hole superradiance (BHSR) from supermassive BH~(SMBH) in M87~\citep{Davoudiasl:2019nlo} and from SMBHs~\citep{Stott:2018opm}. The second line presents the bounds from Lyman-$\alpha$ observations~\citep{Nori:2018pka,Armengaud:2017nkf,Irsic:2017yje,Rogers2020}, from darker to lighter. The bounds from dSphs come from Fornax and Sculptor given in~\citet[][darker]{Gonzales-Morales:2016mkl}, and ~\citet[][lighter]{Marsh:2015wka}, and two constraints with error bars are from Draco and Sextans~\citep{Chen:2016unw}.
    Bellow the dashed line we present the previous and current constraints using UFDs from~\cite{Safarzadeh:2019sre}, and from Eridanus II by~\cite{Marsh:2018zyw}.}
    \label{fig:constraints}
\end{figure}

It is important to understand how the constraints obtained here relate to the bounds obtained from the other independent work. This is shown in Fig.~\ref{fig:constraints}, where we compare our constraint on $m_{\psi}$ from Segue~1, the strongest constraint obtained from this work, shown in red, with a selection of the strongest ones compiled in the review paper~\citep{Ferreira:2020fam}\footnote{There are also other strong constraints not shown here like~\citet{Schutz:2020jox} and \citet{Nadler:2020prv}}. 

The blue shaded regions represent the values of $m_{\psi}$ that are currently excluded by analysis of the correspondent observations. 
In this figure we also show the previous bounds obtained using UFDs from~\cite{Safarzadeh:2019sre} and~\cite{Marsh:2018zyw}. 

We find that the constraint on $m_{\psi}$ for Segue~1 falls in the allowed region of the other observations like CMB, LSS, and BHSR from M87 and SMBHs, even though there is a slight tension with the bound from BHSR from SMBHs.
This result is also compatible with the previous bounds obtained using UFDs, updating on these bounds and obtaining stronger constraints in $m_{\psi}$. 
Our result from Segue 1 is allowed even considering the bounds from Lyman-$\alpha$ measurements, which was to date the strongest bounds requiring $m_{\psi} > 2 \times 10^{-20}\, \mathrm{eV}$, and the first constraint compatible with this bound.
Although not shown in the figure, this is also true for the constraints obtained from the other 17 UFDs, which are all compatible with these bounds within the uncertainty in their mass constraint. 

The Segue~1 constraint seems to be in tension with the ones from the survival of the Eri~II star cluster ("Eridanus II - star cluster" in in Fig.~\ref{fig:constraints}), although compatible with the condition for the existence of a subhalo to host Eri~II, $m_{\psi} > 8 \times 10^{-22} \, \mathrm{eV}$~\citep{Marsh:2018zyw}. 
The survival bound comes from gravitational heating due to the solitonic core. This has recently been challenged~\citep{Schive:2019rrw}, where heating is reduced by tidal disruption of the halo, allowing for the survival of the star cluster for higher $m_{\psi}$ like ours. Future simulations are necessary to verify this effect.

The constraint obtained from Segue~1, however, is in tension with the bounds coming from luminous dSphs, Fornax and Sculptor. These bounds require $m_{\psi}  < 2.9 - 4 \times 10^{-21} \, \mathrm{eV}$~\citep{Gonzales-Morales:2016mkl,Marsh:2015wka}.
Our constraints fall in the excluded region of $m_{\psi}$ from the luminous dSphs, which is also in tension with the bounds from other observations. The same conclusion was found in~\cite{Safarzadeh:2019sre} where a comparative analysis of the constraints on $m_{\psi}$ using data of the half-light radius of dSphs and UFDs was made, using a subgroup of the UFDs using here.  Therefore, $m_{\psi}$ found from the analysis of UFDs cannot explain the density profile of Fornax and Sculptor. This might be indicative that it is challenging for the FDM to address the small-scale problems of the CDM model. This was also pointed out, in the context of the cusp-core problem, using a different argument in~\cite{Burkert:2020laq}.

However, we note that these luminous dSphs could have been affected by baryonic processes. This can change the density structure of their halo, meaning that we are not probing the intrinsic dark matter profile. This might challenge the bounds coming from these systems. 
As discussed in~\cite{Safarzadeh:2019sre}, the FDM soliton profile might be too simplistic, and if allowed to change for different systems, which might also depend on the baryonic effects, would allow a more realistic description of the core of different systems and lead to different bounds on $m_{\psi}$.
In this sense, Milky Way UFDs are ideal objects to probe the nature of DM theory, making us believe that the results presented in this work are a robust measure of the properties of the FDM.

We should also emphasize that our work is based on a fit to a profile for the halo that consists of a soliton core surrounded by an NFW profile, with the relation between the core mass and the halo mass given in~\cite{Schive2014}.
However, different simulations~\citep{Mocz:2017wlg,Mina:2020eik} report a different relation (although the analysis of the dSphs used the same relation as used in this work). 
Another possibility is that the relation between core and soliton size is not as straightforward as the one used in this analysis. These different relations can change the conclusions reached here.
We should also bear in mind that a soliton core in recent FDM simulations is not static but keeps oscillating with amplitude as large as the order of core radius.~\citep{Veltmaat:2018dfz,Schive:2019rrw}. This phenomenon is described theoretically in~\citet{Li:2020ryg}.
This suggests that our analysis based on the simple steady-state modeling of FDM might not describe the FDM halo potential perfectly, which would alter the FDM mass constraints obtained in this work.
Future simulations of the FDM model might help answer these questions. Most importantly, future simulations that include baryons and can test the effect of baryonic effects in the dSph halos would be important to understand this discrepancy between dSphs and other observations.

\section{Conclusions}

In this work, we scrutinized the FDM model through a dynamical analysis of 18 UFDs with the simulation-driven FDM halo density profile. In this work, we were able to put the strongest constraints to date on the FDM mass, with a preference for higher masses $m_\psi \gtrsim 10^{-21} \, \mathrm{eV}$, with the strongest one from Segue 1, $m_\psi=1.1^{+8.3}_{-0.7}\times10^{-19}\ (1\sigma) \, \mathrm{eV}$. This means that the core present in these UFDs is small and shows that for these galaxies, for which the kinematic data available covers a larger spatial extent than the obtained $r_c$, a density profile of a soliton core embedded in an NFW outer profile is needed. The our constraints are compatible with previous bounds from the literature, except classical dSphs~(Fornax and Sculptor), a discrepancy already noted in the literature, which might come from baryonic feedback on the dark matter distribution in these luminous galaxies.
As UFDs are hardly affected by baryonic effects, they are the ideal laboratory to test DM models. With the huge increase in the amount of high quality spectroscopic and astrometric observations in the near future, it will  be possible to use UFDs to improve on the constraints obtained here, showing the power of UFDs to help understand the nature of DM.

\section*{Acknowledgements}
We would like to give special thanks to Eiichiro Komatsu and Simon D. M. White for useful discussions.
This work was supported in part by the MEXT Grant-in-Aid for Scientific Research on Innovative Areas, No.~20H01895~(for K.H.).

\bibliography{manuscript}{}
\bibliographystyle{aasjournal}



\end{document}